\documentclass[acmtog, authorversion, timestamp]{acmart}

\usepackage{physics}
\usepackage{tensor}
\usepackage{derivative}

\usepackage{siunitx}
\usepackage{fmtcount}

\usepackage{fancyhdr}

\usepackage{upgreek}
\usepackage{bbm}
\usepackage{thmtools}
\usepackage{mathtools}
\usepackage{amsmath}
\usepackage{amsfonts}
\usepackage{accents}

\usepackage{verbatim}

\usepackage{dblfloatfix}
\usepackage{subcaption}
\usepackage{xr-hyper}
\usepackage{hyperref}
\usepackage[nameinlink,capitalize]{cleveref}
\crefname{subsection}{Subsection}{Subsections}
\Crefname{subsection}{Subsection}{Subsections}
\crefname{Table}{Table}{Tables}
\Crefname{Table}{Table}{Tables}

\usepackage{tabularray}

\usepackage{xr}
\usepackage{adjustbox}

\def\imgres{_lr}

\newcommand{\inputtikz}[1]{%
    \includegraphics{tikzfigures/#1\imgres}%
}

\usepackage[outline]{contour}

\newcommand\ii{\mathrm{i}}
\newcommand\mpi{\uppi}
\newcommand\ee{\mathrm{e}}

\newcommand\Real{\mathbbm{R}}

\DeclareDocumentCommand\vbar{m}{{ \smash{\widebar{\vb{#1}}} \vphantom{\bar{#1}} }}

\DeclareDocumentCommand\conj{g}{\IfNoValueTF{#1}{\star}{#1^\star}}

\DeclareDocumentCommand\ensemble{s m}{
    \IfBooleanTF{#1}{\langle#2\rangle}{\left\langle#2\right\rangle}
}

\newcommand\R{\vb{R}}
\newcommand\pathR{\vbar{R}}

\newcommand\oprT{\mathcal{T}}
\newcommand\sWDF{\Phi}

\renewcommand\footnotetextcopyrightpermission[1]{} 
\setcopyright{none}
\AtBeginDocument{%
    \addtolength{\footskip}{2.0\baselineskip}%
    \fancyfoot[LE]{\textit{\textbf{Preprint}}}%
    \fancyfoot[RO]{\textit{\textbf{Preprint}}}%
}

\begin{document}

\title{Wave Tracing: Generalizing The Path Integral To Wave Optics}

\author{Shlomi Steinberg}
\email{p@shlomisteinberg.com}
\orcid{0000-0003-2748-4036}
\affiliation{%
    \institution{University of Waterloo}
    \city{Waterloo}
    \country{Canada}
}

\author{Matt Pharr}
\email{matt@pharr.org}
\orcid{0000-0002-0566-8291}
\affiliation{%
    \institution{NVIDIA}
    \city{San Francisco}
    \country{United States}
}


\begin{abstract}

Modeling the wave nature of light and the propagation and diffraction of electromagnetic fields is
crucial for the accurate simulation of many phenomena, yet wave simulations are
significantly more computationally complex than classical ray-based models.
In this work, we start by analyzing the classical path integral formulation
of light transport and rigorously study which wave-optical phenomena can be
reproduced by it.
We then introduce a \emph{bilinear path integral} generalization for wave-optical light transport
that models the wave interference between paths.
This formulation subsumes many existing methods that rely on shooting-bouncing rays or UTD-based diffractions, and serves to give insight
into the challenges of such approaches and the difficulty of sampling good paths in a bilinear setting.

With this foundation, we develop a \emph{weakly-local path integral} based on region-to-region
transport using elliptical cones that allows sampling individual paths that
still model wave effects accurately.
As with the classic path integral form of the light transport equation, our
path integral makes it possible to derive a variety of practical transport algorithms.
We present a complete system for wave tracing with elliptical cones, with applications in light transport for rendering
and efficient simulation of long-wavelength radiation propagation and diffraction in complex environments.

\end{abstract}




\begin{teaserfigure}
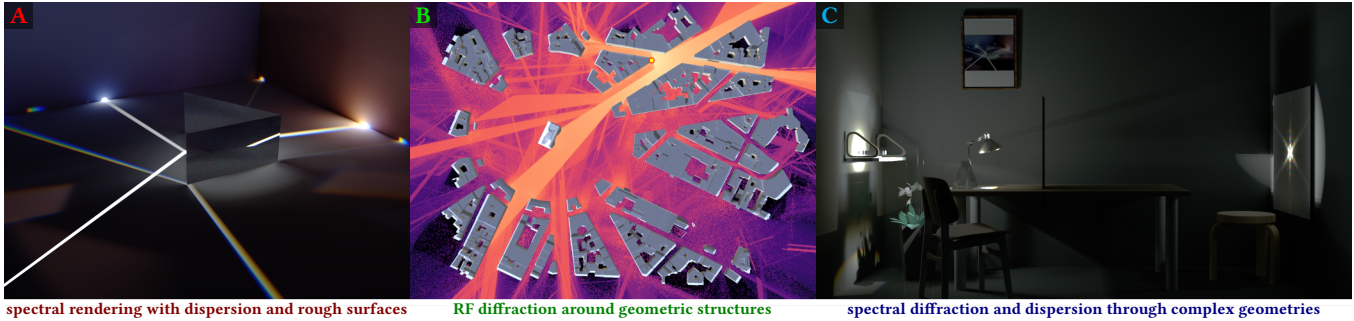

    \centering
    \begin{adjustbox}{width=1\linewidth}
        \inputtikz{teaser}
    \end{adjustbox}
    \vspace*{-5.75mm}
    \caption{
        In this paper we analyze the classical path integral formulation of light transport, and rigorously study what wave-optical phenomena can be reproduced by it.
        We show that some effects, like dispersion and scattering by a restricted class of statistical surface models (rendered in image \textbf{A}), fall under its regime.
        We generalize the classical path integral to a formulation that is able to account for a much wider variety of wave effects, and based on that generalized path integral present a unified framework that is able to:
        (\textbf{B}) simulate long-wave radiation and its propagation and diffraction in complex environments, for example to compute its signal strength (visualized color-coded); and (\textbf{C}) render optical wave effects, such as diffraction by arbitrary geometry.
    }
    \Description{teaser}
    \label{fig:teaser}
\end{teaserfigure}

\maketitle

\thispagestyle{plain}

\section{Introduction}  \label{section_intro}

Ray optics has been the foundation of light transport algorithms in computer graphics since Whitted's introduction of recursive ray tracing.
Especially with the development of path tracing, ray optics light transport has become progressively more sophisticated and has come to be able to model a wide variety of scattering and illumination effects.
Researchers have long sought to extend such ray-based algorithms to model wave effects while maintaining the efficiency of ray models.
One area of focus has been scattering models that account for wave effects when light interacts with surfaces or geometric edges;
another has been the development of algorithms to simulate wave-based light transport.
These areas are complementary in that the light transport depends on accurate scattering and diffraction models
while scattering models require information that is not available in a classical ray-based model: for example, the shape of the electromagnetic wavefronts or optical coherence.
In this work we focus on the light transport problem.

Wave-based models are more computationally expensive than ray-based models, and so we begin by studying a fundamental question:
precisely which wave effects can ray-based light transport accurately simulate, and which can they not?
Under a few assumptions that are characteristic of rendering applications (e.g., sensors that average measurements over long times and a far-field setting),
we rigorously characterize the conditions under which electromagnetic fields obey the Eikonal equation point-wise, and thus, the fields' wavefronts can be propagate using rays.
We further discuss and formally show what classes of materials can be simulated with such light transport models.

With the boundaries of models based on independent rays established, we turn to determining what wave-based models may achieve.
A number of wave simulation methods have been developed that use geometric rays as the underlying transport primitive,
so we start by deriving a \emph{bilinear path integral} that characterizes light transport with interfering ray-based paths.
This work is inspired by Veach's introduction of the path integral formulation of the light transport equation
that transformed the traditional recursive rendering equation into an infinite-dimensional integral over all light carrying paths in a scene~\cite{VeachThesis}.

Our bilinear path integral spans the cases of pairs of paths that do not interact---corresponding to standard path tracing;
pairs of \emph{phase-carrying paths} that always interfere---corresponding to several wave-based light transport models (see \cref{sec:related-work});
as well as pairs of paths that are partially-coherent, including approaches like those based on propagating the Wigner distribution function or the mutual coherence function.
This model clearly shows the difficulties of path sampling with ray-based paths:
although each path individually may have a meaningful contribution, the interaction of a pair of paths may lead to destructive interference and thus, no light transport after all.
Therefore, the efficiency of \emph{local path sampling} is greatly reduced, and path sampling effectively becomes a \emph{global problem}, where all path pairs need to be considered.
While these issues are well-known, our path integral makes the associated sampling challenges clear.

We then turn to develop a \emph{weakly-local path integral} that is based on region-to-region transport.
This model is inspired by \emph{physical light transport} (PLT) \cite{Steinberg_practical_plt_2022,Steinberg_rtplt}, which introduced \emph{weakly-local}---meaning confined to a small spatial region---light transport primitives.
Our formulations generalize their wave-based light transport equation and
maintain the crucial characteristics of modeling transport over small regions of space such that only observable wave effects are accounted for.
As transport is done region-to-region (and not point-to-point), no interference between pairs of paths may arise, making it possible to efficiently construct individual paths via local sampling.
Our formulations further allow the derivation of advanced light transport algorithms for wave models, including a bidirectional model. 

We show that elliptical cones form tight geometric envelopes for PLT's light transport primitives.
Then, based upon the weakly-local path integral formulation, a practical, general-purpose wave transport model is developed.
At its core, our wave transport algorithm can be understood as replacing the classical ray with an elliptical cone.
We discuss how to traverse the scene with elliptical cones, simulate the interaction of their underlying electromagnetic fields with the geometry and materials that fall within an interaction region, and introduce a general importance sampling strategy.
We further show that this algorithm is able to reproduce accurate diffraction effects and do wave simulations for applications in different parts of the electromagnetic spectrum, see \cref{fig:teaser}.

\begin{table}[!b]
\centering
\small
\begin{tblr}{
        colspec = {  l  Q[5.05cm,halign=l]  c  }
    }
    \hline[2pt,purple1]

    \SetCell[c=2]{l} { \textsc{\text{symbol}} } && { \textsc{\text{units}} }
    \\

    \hline
    \SetCell[c=3]{l} { \color{purple!55!black}\emph{paths and their contributions} } &&
    \\
    $\vbar{x}=\vec{x}_1\ldots\vec{x}_n$
    &
    a \emph{path}: a finite sequence of $n$ spatial positions $\vec{x}_j\in\Real^3$
    &
    \\
    $\R$
    &
    a \emph{region}: a bounded, closed subset $\R_j\subset\Real^3$
    &
    \\
    $\pathR=\R_1\ldots\R_n$
    &
    a \emph{weakly-local path}: a finite sequence of $n$ regions (regions may overlap)
    &
    \\
    $f\qty(\vbar{x}), g\qty(\pathR)$
    &
    the \emph{measurement contribution functions} over a path
    &
    {\scriptsize \si{\watt} }
    \\
    $F\qty(\vbar{x},\vbar{y})$
    &
    the \emph{mutual contribution function}, given over a pair of paths $\vbar{x},\vbar{y}$
    &
    {\scriptsize \si{\watt} }
    \\

    \hline[dashed]
    \SetCell[c=3]{l} { \color{violet!55!black}\emph{electrodynamics} } &&
    \\

    $\lambda$
    &
    wavelength of electromagnetic radiation
    &
    {\scriptsize \si{\milli\metre} }
    \\
    $k=\tfrac{2\mpi}{\lambda}$
    &
    wavenumber of electromagnetic radiation
    &
    {\scriptsize \si{\per\milli\metre} }
    \\
    $\vec{k}$
    &
    wavevector
    &
    {\scriptsize \si{\per\milli\metre} }
    \\
    $\psi$
    &
    wavefunction (scalar electric field)
    &
    {\scriptsize \si{\volt\per\metre} }
    \\

    \hline[dashed]
    \SetCell[c=3]{l} { \color{green!30!black}\emph{light transport and Wigner (space-frequency) distributions} } &&
    \\

    $\phi(\vec{r},\vec{k})$
    &
    the Wigner distribution of a \emph{Gaussian beam} (see supplemental \cref{supp:subsection_genrays})
    &
    {\scriptsize \si{\watt} or 1 }
    \\
    $L(\vec{r},\vec{k})$
    &
    Wigner distribution of an emitter's emission
    &
    {\scriptsize \si{\watt} }
    \\
    $W(\vec{r},\vec{k})$
    &
    Wigner distribution of a sensor's sensitivity
    &
    {\scriptsize 1 }
    \\
    $T, \oprT$
    &
    context-dependant interaction or scattering operators acting upon Wigner distributions
    &
    {\scriptsize  }
    \\

    \hline[1pt,purple3]
\end{tblr}
\caption{
    List of symbols and notation in this paper.
}
\label{table:symbols}
\vspace*{-4mm}
\end{table}

\section{Related Work}  \label{sec:related-work}

Our work builds on two classic foundations of ray-based light transport.
First is the path integral formulation of the light transport equation~\cite{veach1997metropolis,VeachThesis}.
This expression of the rendering equation made it clear how to apply non-local sampling techniques to light transport (i.e., that paths do not necessarily need to be sampled incrementally from the sensor) and
has provided a rigorous foundation for deriving more advanced light transport simulation algorithms.
We also build upon the operator expression of light transport developed by \citet{ArvoThesis} and \citet{VeachThesis}.
By expressing light transport and scattering as linear operators on functions describing light emission and sensor response, this formulation allows compact expression of light transport and has enabled analysis of its convergence.

Starting with Stam's pioneering work on modeling diffraction from rough metalic surfaces~\cite{Stam_1999},
there has been considerable work on BSDFs that model wave effects.
Examples include diffraction due to scratches~\cite{Werner2017Scratch,Velinov2018scratches},
iridescent and pearlescent materials~\cite{Guillen:2020:Pearlescence},
thin-film interference~\cite{Huang_Iseringhausen_Kneiphof_Qu_Jiang_Hullin_2020}, and
dispersion via thin dielectric layers above conductors~\cite{Belcour:17,Kneiphof2019}.
Scattering models have also been developed based on measured scattering from diffractive surfaces~\cite{Toisoul2017diffractions},
diffractive surfaces modeled with explicit microgeometry~\cite{Falster2020,Yu_Xia_Walter_Michielssen_Marschner_2023},
and statistical models~\cite{Holzschuch2017,Krywonos2006,Steinberg_speckle_2021}.
Also related is recent work on modeling free space diffraction as a BSDF~\cite{Steinberg_fsd_2024}.

Many techniques have been developed to simulate wave-based light transport.
Approaches based on the Winger distribution function have been developed both in graphics~\cite{Cuypers_Haber_Bekaert_Oh_Raskar_2012} and in optics~\cite{jensen1991methodology,mout2018ray,mackay2021millimetre}.
Outside of graphics, algorithms based on different shooting-bouncing rays methodologies~\cite{weinmann2006ray} or the geometric/uniform theory of diffraction (UTD)~\cite{son1999deterministic,yi2022ray,bilibashi2020dynamic} are often used.
All these approaches rely on ray tracing for transport, and share a fundamental difference compared with the more common ray or path tracing seen in computer graphics:
rays \emph{mutually interfere} with each other.
In this work, we generalize the classical path integral to such \emph{bilinear} transport;
this generalization highlights the additional sampling difficulties that arise with mutually-interfering rays, and in the future may serve as a theoretical foundation for developing better path sampling techniques for such bilinear light transport models.

Recently, physical light transport~\cite{Steinberg_lt_framework_2021,Steinberg_practical_plt_2022,Steinberg_rtplt} has introduced \emph{weakly-local} light transport primitives, which can be understood as tracing beams of light instead of singular rays---enabling the reproduction of a much wider class of wave effects, compared with classical light transport.
Previous implementations of it have approximated the transport over regions with singular rays.
We develop the underlying path tracing theory by generalizing the classical path integral to \emph{region-to-region} transport, and present a solver that models wave transport over volumes using elliptical cones.

\section{What Can Be Simulated With Classical Path Tracing?}  \label{sec:classical_path_integral}

Consider the classical path integral formulation \cite{VeachThesis}:
\begin{equation}
    I = \int_{\Omega} f\qty(\vbar{x}) \dd{\mu\qty(\vbar{x})}
    \label{unitary_path_space_integral}
    ~,
\end{equation}
where $\smash{\vbar{x}\in\Omega}$ is a scene path---connecting an emitter to the sensor over a finite sequence of points---in the space of all paths $\Omega$, and $f$ is the real-valued, non-negative \emph{measurement contribution function}.
The result of $f$ is the power measured by the sensor.

It is well understood that the light transport under the formulation above is governed by the laws of geometric optics.
Nevertheless, some optical phenomena beyond geometric optics may be reproduced accurately by this formalism.
Our objective in this section is then to rigorously answer the question: \emph{to what extent is classical path tracing able to reproduce wave-optical phenomena?}

In our supplemental material, \cref{supp:appendix_geometric_fields}, we formally study the conditions subject to which the wavefronts of an electromagnetic field obey, point-wise, the Eikonal equation, and thereby can be propagated via geometric-optical rays. These conditions can be summarized as: fields (i) of sufficiently high frequency (ii) that propagate very far from their origin (emitter or scattering matter).
We furthermore show that such fields behave locally as plane waves.

We then assume that light is a stochastic ensemble of waves, where every realization conforms to the conditions discussed above.
The wavefronts of such a wave ensemble can be propagated, point-wise, via classical rays as in \cref{unitary_path_space_integral}, and we study what wave phenomena can be reproduced by such a description of light.

\paragraph{Dispersion and color}

The fields admit well-defined temporal and spatial frequencies.
It is possible to model emitters and sensors with arbitrary emission and sensitivity spectra, as well as arbitrary dispersion relations for media, including cross-spectrum re-emission effects, like fluorescence.
These effects can be accurately integrated in a spectral renderer.

One limitation arises:
No interference between spectral samples is permissible, as samples must add up incoherently in \cref{unitary_path_space_integral}.
Therefore, we must assume that sensors time average, and the wave ensemble is limited to the appropriate class of processes (see \cref{supp:sec:classical_pt_analysis_supplemental} in our supplemental material).

\paragraph{Polarization}

The fields also admit well defined directions for their electric components, enabling modeling arbitrary states of polarization, including random polarization (under time-averaging).

\paragraph{Interaction with materials}

Under the assumptions above, we may propagate the wavefronts, point-wise, to an interaction point using geometric optics.
A question arises: what extent of these wavefronts should be considered when simulating their interaction with matter?
That question cannot be answered with the information we have: no additional specifications about the shape of the wavefronts, or their statistics (e.g., spatial coherence) can be deduced, as there is no mechanism in \cref{unitary_path_space_integral} to propagate such information.
Deriving expressions for BSDFs can then be difficult.
We may choose to integrate over very large (or infinite) regions; this would stretch the far field assumption, and also make it impractical to deal with materials that admit explicit scattering features (e.g., a scratched surface, which would require integration over a vast count of scratches).

In \cref{supp:appendix_plane_wave_scattering} of our supplemental material, we show that if we understand the scattering process as a stationary (at least in the wide sense) random process, then the knowledge of the exact integration region is no longer required:
it is enough to assume that this extent is large enough to fully capture the surface's statistics.

Thereby, we may simulate the interaction of light with materials whose scattering is quantified by a statistically wide-sense stationary random process.
The only assumption we make is that the area of integration is large with respect to that process's correlation length.
Such materials include Fresnel reflection and refraction; scattering by rough statistical surfaces (e.g., via the Harvey--Shack formalism), periodic surfaces, or thinly-layered surfaces; or, scattering by a stationary distribution of scatterers in a medium.

\begin{figure*}[t]
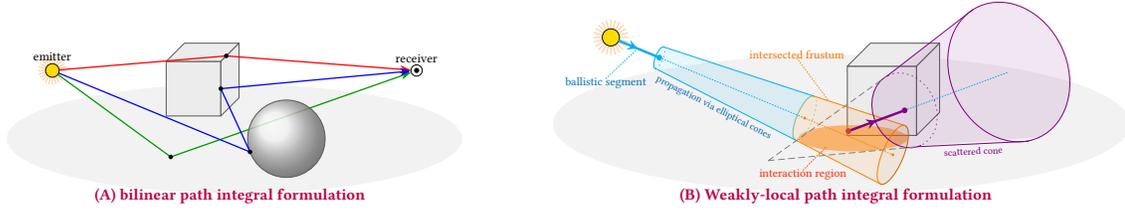

    \centering
    \begin{subfigure}{.45\linewidth}
        \centering
        \begin{adjustbox}{width=.75\linewidth}
            \inputtikz{paths1}
        \end{adjustbox}
        \phantomsubcaption
        \label{fig:paths_bilinear}
    \end{subfigure}%
    \begin{subfigure}{.45\linewidth}
        \centering
        \begin{adjustbox}{width=.95\linewidth}
            \inputtikz{paths2}
        \end{adjustbox}
        \phantomsubcaption
        \label{fig:paths_weakly_local}
    \end{subfigure}%
    \vspace*{-1mm}
    \caption{
        \textbf{Path integrals for wave optics.}
        We develop two generalizations of the classical light transport path integral \cite{VeachThesis}---capable of reproducing a wide range of wave effects.
        (A) The \emph{bilinear} path integral, \cref{section_bilinear_path_integral} (implicitly employed by common methods that use phase-carrying rays, like those that rely on UTD-based diffractions) accounts for the interference between all pairs of paths.
        Bilinearity frustrates local path sampling: assume that a path was sampled; afterwards, another path might be sampled that destructively interferes with the former path, resulting in overall negligible contribution, even with good local path sampling.
        (B) The \emph{weakly-local} path integral, \cref{section_weakly_local_path}, formalizes region-to-region transport.
        In-place of rays, we employ elliptical cones for transport, and formulate the aggregated interaction of light with all materials that fall within a region (illustrated in orange).
        This requires more sophisticated machinery---tracing of elliptical cones and aggregated wave-optical interactions---but allows crucial local path sampling.
        \emph{Ballistic segments} are illustrated using thick lines, and an elliptical conic frustum within which intersected geometry is found in orange (see \cref{sec:wave_tracing}).
    }
    \label{fig:paths}
\end{figure*}
\vspace*{2mm}

\paragraph{Summary}

Following the discussion above, we summarize what forms of optical phenomena are amenable to classical path tracing:
\begin{enumerate}
    \item Light that is an ergodic, wide-sense stationary wave ensemble, of sufficiently large spatial frequencies.
    \item Sensors that time average over periods long with respect to light's temporal coherence.
    \item The far-field assumption is made several times: the points of a sampled path $\smash{\vbar{x}}$ are assumed to be far from each other.
    \item Scattering that is formulated in expectation by statistically wide-sense stationary materials only.
\end{enumerate}
Under these conditions, we prove in our supplemental material that the classical path integral \cref{unitary_path_space_integral} is a sufficient formalism to reproduce all the wave phenomena discussed in this section.

Strictly speaking, it is usually not possible to perfectly satisfy the above set of requirements, for example: at geometric edges or where different geometries with distinct materials intersect (violating assumption 4).
Nevertheless, at optical frequencies, classical path tracing is able to simulate the wave-optical effects discussed above at a reasonable accuracy.

In our supplemental material, we discuss which effects may not be reproduced classically.
As an example that motivates our the derivations in \cref{section_extend_path_integral}, consider a ray passing close to geometry, but not intersecting it.
Under wave optics, part of the energy carried by the ray should have diffracted around the geometry.
However, we may not detect this situation with classical ray tracing (a ``does this ray travel close to geometry'' query would be needed), so the energy that is propagated straight on is overestimated.
To amend the error, one may consider two approaches: we could (i) allow other paths to fix these erroneous contributions, by facilitating interference between different paths, as is done with ray-based techniques such as UTD; or, (ii) replace ray queries with a volumetric query, in order to sample \emph{regions}---and not singular points---where wave-optical interactions occur.
In the following section we generalize the path integral to both of these approaches.

\section{Generalizing the Path Integral Formulation}  \label{section_extend_path_integral}

\subsection{Bilinear Path Integral}  \label{section_bilinear_path_integral}

We extend the classical path integral with a map
$\smash{ F : \Omega\cross\Omega \to \Real }$,
the \emph{mutual contribution function}, producing a \emph{bilinear path integral}:
\begin{equation}
    I =
        \int_{\Omega\cross\Omega}
            F\qty(\vbar{x},\vbar{y}) 
            \dd{\mu\qty(\vbar{x})} 
            \dd{\mu\qty(\vbar{y})}
    \label{bilinear_path_space_integral}
    ~.
\end{equation}
We only consider maps $F$ that fulfill the following properties:
\begin{enumerate}
    \item symmetry, $F(\vbar{x},\vbar{y})=F(\vbar{y},\vbar{x})$;
    \item non-negative contribution over a single path, $F(\vbar{x},\vbar{x})\geq0$;
    \item Cauchy--Schwarz inequality, $\abs*{F(\vbar{x},\vbar{y})}^2 \leq F(\vbar{x},\vbar{x})F(\vbar{y},\vbar{y})$.
\end{enumerate}
Note, when $\vbar{x}\neq\vbar{y}$, the map $F(\vbar{x},\vbar{y})$ may take negative values.
It is easy to see that the properties above mandate that the path integral in \cref{bilinear_path_space_integral} integrates to a non-negative $I$, as desired (see \cref{supp:supp_subsection_nonnegativity_bilinear} in the supplemental for a proof).

\paragraph{As a statistical distribution}

To see how the mutual contribution function $F$ connects to optics, 
let $\psi$ be some realization of a statistical wave ensemble (i.e., $\psi$ is some electric field).
Given a geometric path $\vbar{x}$, we will let $\psi(\vbar{x})$ denote the field strength that is transported over that path.
(We briefly mention a few formalisms that enable such field transport over a path later).
Then, the measurement in expectation becomes:
\begin{align}
    I =& 
        \ensemble{
            \abs{\int_{\Omega} \psi\qty(\vbar{x}) \dd{\mu\qty(\vbar{x})}}^2
        }
    =
        \int_{\Omega\cross\Omega}
            \ensemble{ \psi\qty(\vbar{x}) \conj{\psi}\qty(\vbar{y}) }
            \dd{\mu\qty(\vbar{x})}
            \dd{\mu\qty(\vbar{y})}
    \label{stochastic_path_space_integral}
    ~,
\end{align}
where $\conj{\psi}$ is the complex conjugate of $\psi$, and we formally interchanged the order of ensemble averaging and integration.
The above takes an identical form to \cref{bilinear_path_space_integral}, with 
$\smash{ F(\vbar{x},\vbar{y}) = \ensemble*{ \psi(\vbar{x}) \conj{\psi}(\vbar{y}) } }$
quantifying the ensemble's second-order statistics (its cross-correlation), and is the \emph{mutual coherence function} in the optical context.

The statistical formulation above may also account for temporal fluctuations, e.g., when $F$ operates on paths with different wavelengths (under spectral rendering).
This arises in practice in simulations where sensors do not time average (or averaging times are too short to induce temporal decoherence), for example for acoustic simulations with observable interference across spectral samples.
For ergodic ensembles $\psi$, the ensemble averages can indeed be understood as time averages, though we keep the formulation general.

\paragraph{Examples}

In the limiting case where the mutual contribution function is a proportional to a Dirac delta,
$
    F(\vbar{x},\vbar{y}) =
        \abs*{\psi(\vbar{x})}^2 
        \delta(\vbar{x},\vbar{y})
$~,
the above reduces to the classical unitary case, with 
$\smash{ f=\abs*{\psi(\vbar{x})}^2 }$ 
being the classical measurement contribution function.
The Dirac delta formalizes the fact that all paths are assumed to be perfectly mutually uncorrelated---\emph{incoherent}---and no interference arises.
Therefore, the bilinear path integral generalizes the classical path integral.

Another important special case is path tracing with \emph{phase-carrying rays}, where every pair of paths are fully coherent in their measurement and interfere.
Let the map $F$ take the form:
\begin{align}
    F\qty(\vbar{x},\vbar{y}) &=
        \abs{\psi\qty(\vbar{x})} 
        \abs{\psi\qty(\vbar{y})}
        \Re \ee^{-\ii k \qty(\abs*{\vbar{x}}-\abs*{\vbar{y}})}
    ~,
\end{align}
where $\abs*{\vbar{x}}$ is the path's total distance (in world units), 
and $\smash{ k=\tfrac{2\mpi}{\lambda} }$ is the wavenumber with wavelength $\lambda$.
Clearly $F$ fulfills the desired properties of the mutual contribution function.
The terms $k\abs*{\vbar{x}}$ are then the phases carried by each ray, and the complex exponent term quantifies the mutual interference between every pair of paths.
This formalism of phase-carrying rays encompasses several common light transport methods, including UTD-based simulations---very common in acoustics and RF simulations---and other methods that rely on mutually-interfering phase-carrying rays.

In between the above two extremes---perfect incoherence and perfect coherence---partially-coherent formalisms arise.
Such formalisms include light transport formalisms that employ the Wigner distribution function or the mutual coherence function directly.

\paragraph{Local path sampling}

Local path sampling \cite{Kajiya_1986} is a very common technique to construct complete paths $\vbar{x}$, where each point in the path is recursively sampled \emph{locally} by sampling the local interaction function and then ray tracing.
Practical light transport often relies on the ability to perform good local path sampling.

With bilinear light transport, local path sampling is hampered by the fact that the mutual contribution $F(\vbar{x},\vbar{y})$ usually cannot be evaluated without full knowledge of the paths $\vbar{x},\vbar{y}$, and may annihilate the paths' contributions.
That is
\begin{align}
    \ensemble{
        \abs{\psi\qty(\vbar{x}) + \psi\qty(\vbar{y})}^2
    }
        =&
        \ensemble{\abs{\psi\qty(\vbar{x})}^2}
        +
        \ensemble{\abs{\psi\qty(\vbar{y})}^2}
        +
        2F\qty(\vbar{x},\vbar{y})
\end{align}
may vanish entirely.
That is, even if a path $\vbar{x}$ was constructed with effective local path sampling, another path $\vbar{y}$ might annihilate its contribution (illustrated in \cref{fig:paths_bilinear}).
Under the bilinear setting, \emph{path sampling always becomes a global problem}.
The classical path integral, \cref{unitary_path_space_integral}, does not suffer from this issue, as its integrand is always non-negative.
It is also noteworthy that constructive/destructive interference happens at a high frequency across the scene, making techniques like path guiding not effective.
The difficulty with local path construction is a fundamental challenge in path tracing simulations that employ typical UTD, or other phase-carrying rays, formalisms.

Other forms of non-local path construction, like Metropolis light transport, could be applied to the bilinear path integral.
We leave this to be explored in future work.

\subsection{Weakly-Local Path Integral}  \label{section_weakly_local_path}

The classical path integral, \cref{unitary_path_space_integral}, formalizes point-to-point light transport.
We now generalize that formulation to region-to-region light transport (illustrated in \cref{fig:paths_weakly_local}):
\begin{equation}
    I = \int_{\Omega} g\qty(\pathR) \dd{\mu\qty(\pathR)}
    \label{weakly_local_path_space_integral}
    ~,
\end{equation}
where $\smash{\pathR=\R_0\R_1\ldots \R_n}$ is a generalized path defined as sequence of bounded regions $\R_j\subset\Real^3$, 
$\Omega$ is now the set of all such sequences of all bounded subsets of space, and $\mu$ is then the appropriate product measure:
$
    \mu(\pathR) = \mu(\R_0) \ldots \mu(\R_n)
$.
$g$ is the path contribution function, which can be written in operator notation:
\begin{align}
    \!\!\!
    g\qty(\pathR) =
        \smash{\Big<}
            W
            ,\;
            T_{\R_{n-2} \to \R_{n-1} \to \R_n}
            &T_{\R_{n-3} \to \R_{n-2} \to \R_{n-1}}
            \ldots
            \nonumber \\
            &T_{\R_0 \to \R_1 \to \R_2}
            L_{\R_0\to \R_1}\qty(\R_0)
        \smash{\Big>}
    \label{weakly_local_path_space_contribution}
    ~.
\end{align}
$T$ are generic transport operators, generalizing the classical BSDF to region-to-region transport.
The emission function $L$ sources a distribution of light from a given initial region $\R_0$:
\begin{align}
    \phi_1 = L_{\R_0\to \R_1}(\R_0)
    ~.
\end{align}
The transport operators then transform these distributions, as follows, thereby propagating them from region to region:
\begin{align}
    \phi_{j+1} = T_{\R_{j-1} \to \R_{j} \to \R_{j+1}} \phi_j
    ~.
\end{align}
Finally, the inner product of the sensor sensitivity function $W$ over the distribution of light quantifies the sensor's response to a distribution and yields the measurement contribution of the path $\pathR$:
\begin{align}
    g\qty(\pathR) =&
        \expval{
            W, 
            \phi_n
        }
    \label{weakly_local_path_contribution_integration}
    ~,
\end{align}
which must be real and non negative, $\smash{g\qty(\pathR)}\geq0$.
We discuss the above in additional detail and prove the non-negativity of \cref{weakly_local_path_contribution_integration} in our supplemental material.

In the following section, we understand the distributions $\phi$ as a specific class of space-frequency (Wigner) distributions, 
and the transport operators $T$ describe the interaction of these fields with matter in a region $\R_j$ and their propagation to the next region.
In general, $\phi$ can take other forms:
for example, they may be correlation functions, quantifying a wave ensemble's second order statistics, or realizations from a wave ensemble.
We keep the definition abstract to make it applicable to different light transport formalisms.

Our weakly-local formulation of the path integral makes explicit the regions $\R_j$ over which light transport takes place.
It is easy to see that \cref{weakly_local_path_space_integral} generalizes the classical formulation: when the regions are reduced to points, $T$ to classical BSDFs, and $\phi_j$ to radiance point samples, the above reduces to the classical point-to-point transport.
However, \cref{weakly_local_path_space_integral} is also more powerful than the classical formulation:
we now have a built-in mechanism to quantify the shapes of the wavefronts of the underlying electromagnetic field (via $\phi_j$),
and to quantify the regions over which weakly-local interactions of light with matter may take place (via $\R_j$).

\paragraph{Local path sampling}

Unlike the bilinear generalization in \cref{section_bilinear_path_integral}, the contribution $g$ is crucially always non negative---no interference across samples may arise.
Therefore, local path sampling may proceed in a manner essentially identically to the classical case: sample an initial region $\R_0$, and then subsequent regions, using the emission function $L$ and the transport operators $T$, respectively.

\section{Weakly-Local Light Transport}  \label{section_wo}

The ability to construct paths via local path sampling with our weakly-local path integral formulation, discussed in \cref{section_weakly_local_path}, is a major advantage over its bilinear counterpart.
It comes at the cost of increased analytic and algorithmic complexities of the simulation:
light transport is now formulated in terms of regions and not points, and our intersection queries are no longer ray--geometry intersections but volumetric queries.
In this section we address these difficulties and present a general weakly-local light transport framework targeting wave-optics simulations.

We build upon the physical light transport (PLT) framework for wave-optics simulation \cite{Steinberg_rtplt,Steinberg_practical_plt_2022}.
Under that framework, the emission and sensing distributions of a rather general class of emitters and sensors can be written as a (countable) incoherent superposition of \emph{Gaussian beams}.
For example, the space-frequency (Wigner) distribution of an emitter is decomposed as
\begin{align}
    L\qty(\va{r},\va{k})
        =&
            \sum\nolimits_j \phi_j\qty(\va{r},\va{k})
    \label{plt}
    ~,
\end{align}
where we use $\smash{\va{r}}$ for spatial coordinates, and $\smash{\va{k}}$ is a wavevector---direction of propagation scaled by the wavenumber $\smash{k=\tfrac{2\mpi}{\lambda}}$---and $\smash{\phi_j}$ are space-frequency Gaussian distributions.
A similar decomposition can be written for a sensor's sensitivity distribution.
Please see our supplemental material for an extended overview of the relevant parts of PLT and explicit expressions for $\phi_j$.

Path tracing and rendering with PLT involves: sourcing a sample, i.e. a beam $\phi_j$, from an emitter (or a sensor); propagating it across the scene and simulating its interaction with matter; and finally integrating over a sensor (or an emitter).

\subsection{Wave Tracing}  \label{sec:wave_tracing}

A fundamental difference with classical path tracing is that the distributions $\phi_j$ are not point samples, but \emph{weakly-local} beams, confined to a bounded spatial region.
We show in our supplemental material that the spatial extent of $\phi_j$, as it propagates in unobstructed space, traces an \emph{elliptical cone}.
Therefore, these elliptical cones form tight \emph{geometric envelopes} around these beams, and their geometry is defined by the spatial and wavevector means and variances of $\phi_j$.
As the Gaussian beam is fully described by its first two moments, the geometric envelope defines the Gaussian beam $\phi_j$, and vice versa.

Tracing the elliptical conic envelopes through space, i.e. \emph{wave tracing}, serves to construct the path $\pathR=\R_1\R_2,\ldots,\R_n$, composed of the weakly-local regions over which light transport takes place.
Evaluating the path contribution function $g(\pathR)$ (\cref{weakly_local_path_space_contribution}) over that path yields the sample contribution.
In our renderer, each beam is parametrized by:
\begin{enumerate}
    \item mean wavelength $\lambda_0$;
    \item Stokes parameters quantifying its polarization and power;
    \item its elliptical conic envelope, parametrized by
        (i) mean spatial position $\va{x}_0$ at its origin,
        (ii) mean direction of propagation $\vu{d}$,
        (iii) major and minor axes $\smash{\va{a},\va{b}}$, of lengths $a=\abs*{\va{a}},b=\abs*{\va{b}}$,
            and
        (vi) major and minor opening (half) angles $\alpha_a,\alpha_b\geq0$.
\end{enumerate}
Sourcing a beam is identical to PLT, and we provide explicit expressions in our supplemental material.

\paragraph{Traversal}

On free-space propagation of distance $z\geq0$, the geometry of the elliptical cone is transformed as follows:
\begin{align}
    \!\!\!
    \smash{
        \va{x}_0 \to \va{x}_0 + z\vu{d} ~,
    }
    &&
    a \to a + z\tan\alpha_a
    && \qand* && \!\!\!\!
    b \to b + z\tan\alpha_b
    \label{cone_propagation}
    ~,
\end{align}
with the other parameters unchanged.
We employ a BVH for elliptical cone traversal acceleration \cite{Umut_cone_tracing_2025}.
Traversal is similar to a ray-based traversal, though a fundamental difference is that a cone may intersect multiple triangles.
As we traverse the BVH we collect intersected triangles, and keep track of the distance to the closest triangle $d_\textrm{min}$. The traversal routines returns all triangles that intersect the elliptical cone within a distance range $[d_\textrm{min}, d_\textrm{min}+\delta)$, where $\delta$ controls the depth of the intersection range.
The ``intersection distance'' is defined as the propagation distance along the mean $\vu{d}$ (and not the radial distance), therefore the computed intersection region is an \emph{elliptical conic frustum} (illustrated in orange in \cref{fig:paths_weakly_local}).
We set ad hoc $\delta=2a$, i.e., twice the length of the major axis, which is sufficient for most applications and a good compromise between accuracy and performance for others.

Once the elliptical conic frustum that contains all the intersected triangles has been computed, it becomes the next interaction region $\R_{j+1}$.
The wave tracing process then repeats recursively in order to sample the complete path $\pathR$:
the interaction of the beam with the geometry at $\R_{j+1}$ is simulated and a scattered beam---and its elliptical conic envelope---is sampled (discussed in \cref{section_diffractions}), which is then beam traced to compute the next interaction region.

\paragraph{Ballistic propagation}

In order to allow a scattered beam to propagate away from an interaction region (without extensive self-intersections), as well as to allow it to explore tight regions (e.g., a waveguide that might be narrower than the geometric envelope), we perform \emph{ballistic propagation}~\cite{Lemieux1998-aj}.
Immediately after sourcing or interaction, we assume the beam takes a \emph{ballistic path}---traces a ray with a vanishing cross section---for a short distance, as follows:
\begin{enumerate}
    \item We propagate the beam as a ballistic particle---via ray tracing---for a segment of up to $B$ wavelengths in distance.
    \item If the ray intersects geometry, a normal interaction is performed; after which we return to step 1.
    \item Otherwise, we attempt to resume cone tracing from the end of the ballistic segment:
    \begin{enumerate}
        \item if the elliptical cone immediately (without any further propagation) intersects geometry:
        \begin{enumerate}
            \item $B$ is doubled, up to some predefined maximum $B_\textrm{max}$;
            \item return to step 1.
        \end{enumerate}
        \item otherwise, elliptical cone tracing is continued.
    \end{enumerate}
\end{enumerate}
The doubling of the ballistic segment length $B$ is designed to avoid excessive attempts to terminate ballistic propagation.
Ballistic propagation also alleviates some of the chronic pains of cone tracing, like a beam propagating parallel to a wall and repeatedly intersecting it.
A ballistic path is illustrated via a thick line in \cref{fig:paths_weakly_local}.

In addition to enabling a beam to explore tight confines in a physically-sensible manner, ballistic propagation also serves as a major performance optimization: 
the most expensive part of cone tracing is when the cone passes close to geometry and many expensive cone--primitive intersection tests are required.
Our renderer sets ad hoc the ballistic segment lengths $B$ and $B_\textrm{max}$ to \num{10} and $2^{10}$ wavelengths, respectively.

\begin{figure*}[tb]
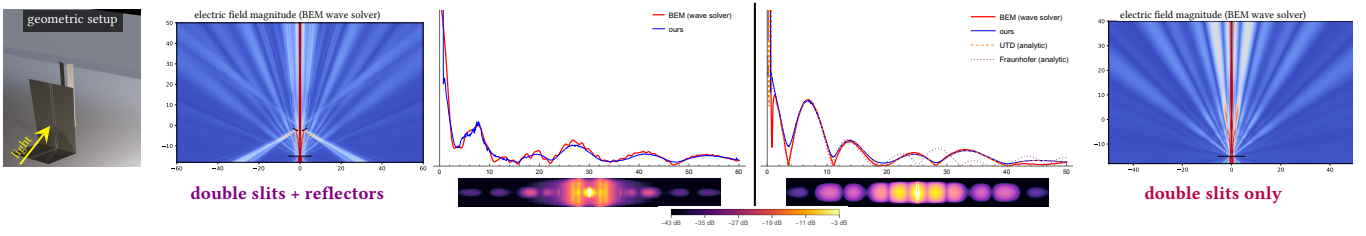

    \centering
    \begin{adjustbox}{width=1\linewidth}
        \inputtikz{wavesolver}
    \end{adjustbox}
    \vspace*{-5mm}
    \caption{
        \textbf{Comparison with a wave solver:}
        a collimated beam impinges upon a screen with two slits cut in it, and a couple of conductive reflectors are placed after the screen (setup illustrated on the left).
        Diffracted light reaches the far wall and the resulting diffraction pattern is rendered.
        We use a BEM wave solver \cite{bempp} to solve for the electric field over a 2D vertical slice, and compare with results computed using our renderer (UTD free-space diffractions), see plots.
        Our results are shown below the plots.
        Multi-edge diffractions happen over two regions---slits and then reflectors---before reaching the wall.
        We also perform the experiment without the reflectors (double slits only) and plot analytic solutions to Fraunhofer and UTD diffractions: as expected our results closely match UTD and are more accurate than Fraunhofer diffraction.
        (Run costs) ours (full 3D): \textbf{4 min} and \SI{195}{\mega\byte} of memory, BEM (2D slice): \textbf{2 hours, 27 min} and \SI{60}{\giga\byte}.
    }
    \label{fig:wave_solver}
\end{figure*}
\vspace*{2mm}

\subsection{Interactions and Diffractions}  \label{section_diffractions}

Let $\R_j$ be an interaction region: that is, the intersection of an elliptical conic frustum with geometry.
The incident beam $\phi_j$ interacts with the matter and materials, and the scattered distribution from an interfering superposition of $N\geq1$ triangles (and their materials) is a \emph{bilinear} combination that accounts for interference:
\begin{align}
    \sWDF\qty(\va{r},\va{k})
        =& \sum\nolimits_{l=1}^N
            \oprT_l \phi_j +
            2\sum\nolimits_{m>l}^N
                \oprT_{lm} \phi_j
    \label{interference}
    ~.
\end{align}
The terms $\oprT_l$ are operators which compute the scattered light produced by each triangle (or material) $l$.
The double-subscripted scattering operators $\oprT_{lm}$ quantify the bilinear interference terms between all triangle pairs.
This is a very general expression---any interfering wave-optical interaction needs to account for such interference terms.
For example, if a triangle's material models a rough surface, then $\oprT_l$ computes the scattering of a Gaussian beam by that surface over the triangle's extent; often, an approximate surface scattering model will be used.
See our supplemental material for more details.

Our goal is to sample a new beam and its envelope from \cref{interference}, enabling us to beam trace it and compute the next interaction region $\R_{j+1}$.
Formally, the power contained in some scattered Gaussian beam $\phi^\prime$ is the inner functional product:
\begin{align}\!
    I^\prime 
        =& \expval{\phi^\prime,\sWDF}
        = \sum\nolimits_{l=1}^N
            \expval{\phi^\prime,\oprT_l \phi_j} +
            2\sum\nolimits_{m>l}^N
                \expval{\phi^\prime,\oprT_{lm} \phi_j}
    \label{phi_power}
    ~,
\end{align}
with the functional product taken over the space-frequency (Wigner) space:
$\expval{f,g}=\!\int\!\dd{\va{r}} \smash{\dd{\va{k}}} f\conj{g}$.
The (unknown) beam $\phi^\prime$ defines its geometric envelope, which in turn defines the next interaction region.
Therefore, if we could derive from \cref{phi_power} and expression for $I^\prime$ as a function of the envelope's geometry, we could importance     sample a scattered beam from the interaction.

PLT invariants help us narrow down what kind of $\phi^\prime$ we should consider:
The space-frequency bandwidth of the Gaussian beams $\phi^\prime$ that decompose the scattered light $\sWDF$ cannot decrease (we may not ``break a beam into smaller beams'').
Because the Gaussian beam is fully described by its elliptical conic envelope, this constraint implies a geometric invariant: the envelope of the new $\phi^\prime$ contains the current interaction region $\R_j$ and propagates into an identical solid angle as the incident beam.

To satisfy this geometric invariant, when constructing a new elliptical conic envelope for an arbitrary new tracing direction $\smash{\vu{d}^\prime}$ we proceed as follows:
we solve for the elliptical cone that contains $\R_j$ and propagates into the same solid angle as the incident beam, i.e. $\tan\alpha_a^\prime\tan\alpha_b^\prime=\tan\alpha_a\tan\alpha_b$.
The ratio $\alpha^\prime_a/\alpha^\prime_b$ is chosen such that the beam's eccentricity remains constant (this is valid and simplifies some intersection tests).
This fixes the envelope, which in turn defines the scattered Gaussian beam $\phi^\prime$.

In general, computing closed forms expressions for the inner products in \cref{phi_power} can be difficult.
In our supplemental material, \cref{supp:sec:wave_tracing_supplemental}, we study what assumptions (or approximations) need to be imposed upon the scattering operators $\oprT_l$, in order to be able to write these inner products as functions of only incident and scattering directions and interaction footprint.
We call such materials \emph{Fraunhofer materials}, and show that for these materials the inner products above take an analytically-simpler form.
We also show that classical BSDFs form a strict subset of Fraunhofer materials.

\subsubsection{Importance sampling} \label{section_importance_sampling}

We focus on the task of choosing the new direction $\smash{\vu{d}^\prime}$, and present a novel, general importance sampling strategy for interfering interactions.
Our aim is to importance sample \cref{phi_power}.
In general, energy conservation mandates the following:
\begin{align}
    \abs{\expval{\phi^\prime,\oprT_{lm}\qty(\phi_j)}}^2 
    \leq
        \expval{\phi^\prime,\oprT_l\qty(\phi_j)}
        \expval{\phi^\prime,\oprT_m\qty(\phi_j)}
    ~.
\end{align}
By the above, the terms $\expval{\phi^\prime,\oprT_l\qty(\phi_j)}$ may act as a proposal distribution for the entire sum in \cref{phi_power}, and it must hold that
\begin{align}
    I^\prime \leq N
        \sum\nolimits_{l=1}^N
            \expval{\phi^\prime,\oprT_l\qty(\phi_j)}
    ~.
\end{align}
This is a classical setting for rejection sampling.

Assume that we are able to importance sample a new direction from each of the terms $\expval{\phi^\prime,\oprT_l\qty(\phi_j)}$, but not from the more complicated interference terms.
Then, our importance sampling strategy involves rejection sampling the interfering sum in \cref{phi_power}:
\begin{enumerate}
    \item Draw a direction $\smash{\vu{d}^\prime}$ by importance sampling the (incoherent) sum of non-negative terms
          $\sum_l\expval{\phi^\prime,\oprT_l\qty(\phi_j)}$.
    \item For the sampled $\smash{\vu{d}^\prime}$, evaluate the actual power $I^\prime$ using \cref{phi_power}, and incoherent power $\widetilde{I}=\sum_l\expval{\phi^\prime,\oprT_l\qty(\phi_j)}$.
    \item Rejection sample:
    \begin{enumerate}
        \item draw uniformly distributed $u\in[0,1)$;
        \item if $u < I^\prime / (N \widetilde{I} )$ accept $\smash{\vu{d}^\prime}$, otherwise return to step 1.
    \end{enumerate}
\end{enumerate}
The above produces perfectly-distributed samples, and the expected number of trials is $N$ \cite{Casella2004}.

\subsubsection{Free-Space Diffractions}  \label{section_fsd}

When an elliptical cone is partially occluded by geometry, one of the scattering operators $\oprT_l$ in \cref{interference} is a free-space propagation operator, and its restriction to a part of the incident beam gives rise to \emph{free-space diffractions}.
We consider two methods to simulate free-space diffractions:
(i) building directly upon the Fraunhofer edge-based diffraction method of \citet{Steinberg_fsd_2024}; and,
(ii) by utilizing UTD.
Both methods ultimately reduce to writing the diffracted term as an interfering sum, in the manner of \cref{interference}, therefore free-space diffraction operators should be understood as a special case of it.
We summarize each method.

\paragraph{Fraunhofer edge-based diffraction}

\begin{figure*}[tb]
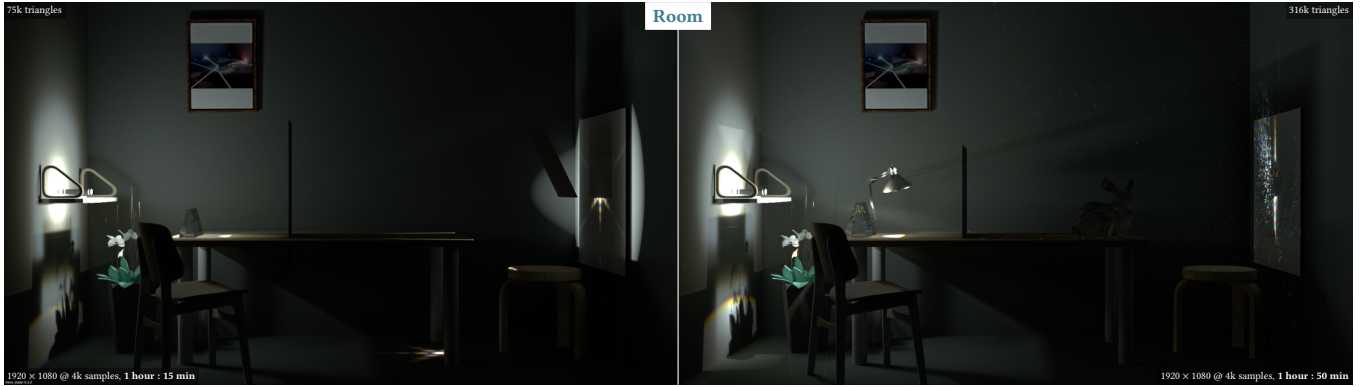

    \centering
    \begin{adjustbox}{width=1\linewidth}
        \inputtikz{results_room}
    \end{adjustbox}%
    \vspace*{-0.5mm}
    \caption{
        \textbf{Wave optics rendering:}
        A metal screen is placed between a powerful collimated light source and a wall.
        A geometric aperture is cut in the middle of the screen: (left) star shaped or (right) vertical double slits.
        Our simulation reproduces the expected diffraction patterns.
        To make the light transport more challenging we place a (left) mirror or (right) dielectric bunny, making direct sampling techniques difficult.
        This shows that we able to importance sample the entire multi-edge diffraction well.
        This scene is a miniaturized scene: the length of the far wall is \SI{8}{\centi\metre}.
        The cross-sectional areas of the elliptical conic envelopes on interactions are \SI{.2(1.0)}{\square\milli\metre} (mean and standard deviation), and intersect with \num{14(261)} triangles.
        Fraunhofer edge-based free-space diffractions are used (see \cref{section_fsd}).
    }
    \label{fig:render_room}
\end{figure*}

This method proceeds exactly as in \citet{Steinberg_fsd_2024}: the edges of the silhouette of the diffracting aperture in an interaction region $\R_j$ are extracted.
Each edge $e_l$ gives rise to an edge diffraction operator $\oprT_{e_l}$, while interference operators $\oprT_{e_le_m}$ account for the interference between the edges.
Together, these operators shape the Fraunhofer diffraction pattern.
\citet{Steinberg_fsd_2024} provide explicit formulae for both, as well as a simple importance sampling strategy for the edge diffraction operator $\oprT_{e_l}$, but not for the more complicated interference operators $\oprT_{e_le_m}$.
Our importance sampling strategy from \cref{section_importance_sampling} improves upon theirs by importance sampling the entire interfering sum of edge diffractions.
This enables high quality sampling, and we show in \cref{fig:render_room} that we are able to render complex diffraction patterns under complex light transport.

\begin{figure*}[tb]
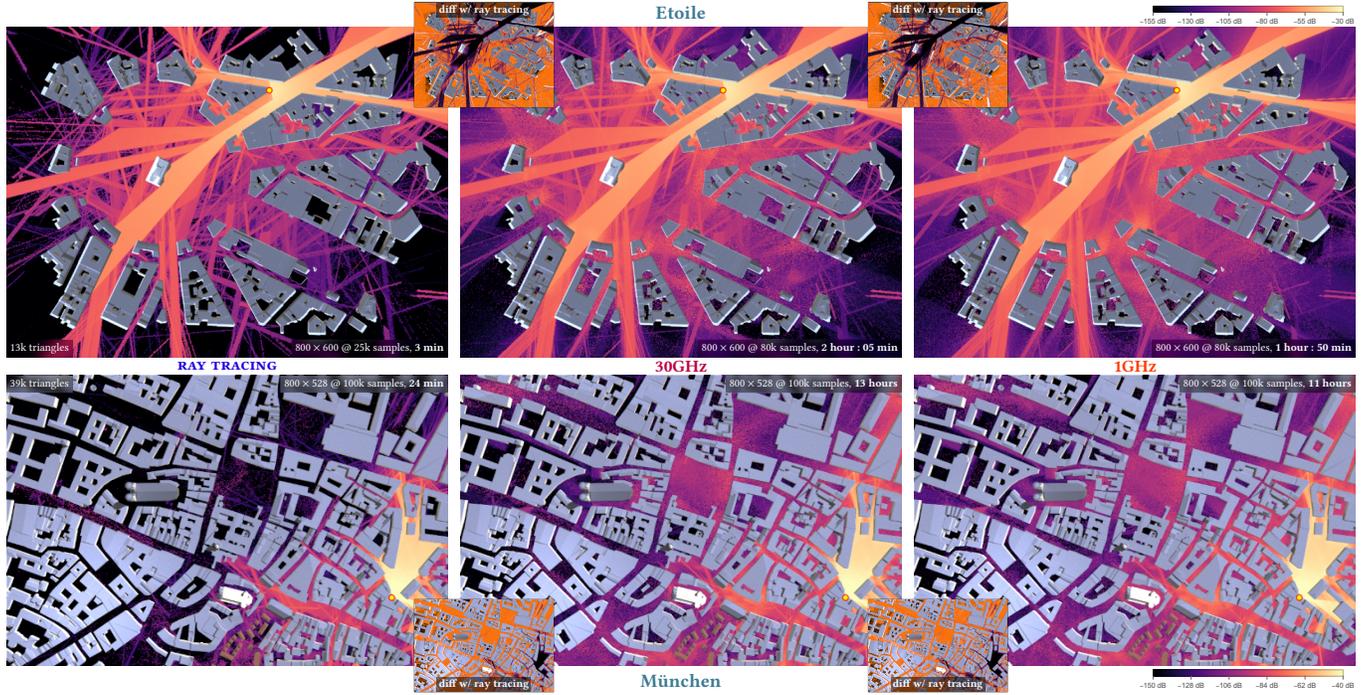

    \centering
    \begin{adjustbox}{width=1\linewidth}
        \inputtikz{results_coverage}
    \end{adjustbox}%
    \vspace*{-2.5mm}
    \caption{
        \textbf{Signal coverage simulations} in realistic city scenes (from \citet{sionna}), done in full 3D.
        Visualized is the received signal strength (rss) in decibels that reaches the street level.
        Emitter position is marked with a red-yellow circle.
        Simulation is done using (left) pure ray tracing with no diffractions; or with UTD-based diffractions using a (middle) 30GHz or (right) 1GHz carrier.
        Note the greater rss reaching shadowed streets and areas when diffractions are simulated.
        Buildings' surfaces are perfectly-specular surface with wavelength-dependent refractive indices matching common construction materials \cite{ITUR_P.2040_3}
        Coverage is computed over ground areas of (Etoile) \SI{.84}{\kilo\metre}, (M\"unchen) \SI{1}{\kilo\metre} in width.
        The elliptical conic envelopes' cross-section areas and primitive counts on interactions (means and standard deviations):
        (Etoile 30GHz) \SI{56(250)}{\square\metre}, \num{7(21)};
        (Etoile 1GHz) \SI{90(290)}{\square\metre}, \num{9(22)};
        (M\"unchen 30GHz) \SI{49(298)}{\square\metre}, \num{11(129)};
        (M\"unchen 1GHz) \SI{84(364)}{\square\metre}, \num{10(107)}.
        UTD-based free-space diffractions are used (see \cref{section_fsd}).
    }
    \label{fig:render_signal_coverage}
\end{figure*}%

\paragraph{UTD-based diffraction}

Fraunhofer-based diffractions are less accurate for long-wavelength radiation, assume perfect conductors, ignore wedge geometry, and it can be difficult to correctly classify edges which are partially occluded by triangles within the interaction region.
As an alternative, we propose a method that relies on the uniform theory of diffraction (UTD) \cite{McNamara1990-ez} to simulate free-space diffractions.
This method can be understood as tracing a ray bundle confined within the beam's envelope.
At each interaction region $\R_j$, we attempt to connect rays in the bundle from the previous region $\R_{j-1}$ to the next one $\R_{j+1}$ over edges in $\R_j$, by checking if the diffracted Keller cone for each edge in $\R_{j}$ falls upon the destination.

Similar to before, the rays in the bundle give rise to diffracted terms $\oprT_{e_n}$, however these are not Fraunhofer materials, and importance sampling is more difficult.
Our sampling approach is very basic: a conservative frequency-dependent Gaussian is fitted to the UTD edge diffraction function, and at each interaction region an edge (that is involved in the diffraction) is selected at random and the next direction is sampled from that Gaussian.
We do not consider the interference terms when sampling, nor perform multiple importance sampling (MIS) over the different edges.
This leads to rather poor sampling; a sophisticated importance sampling strategy for UTD-based ray bundle diffractions is left for future work.

\section{Discussion}   \label{sec:discussion}

\cref{sec:wave_tracing,section_diffractions} describe the primary components of wave tracing as done in our system.
We construct region-to-region paths by tracing the elliptical conic envelopes of Gaussian beam samples.
Cone tracing allows us to sample a complete interaction region, with all the contained geometry that the Gaussian beam interacts with.
In contrast to ray-based frameworks like Sionna \cite{sionna}, we do not need to perform a shooting-bouncing rays exploration pass in order to detect edges (which might fail to detect relevant edges that may lie in shadowed, unexplored regions, or due to high tesselation of the geometry).
Cones are inherently able to sample zero-measure features, like the edges needed for free-space diffraction, and could also be used to sample caustics in a unidirectional path tracer; to be explored in future work.

The beams and their envelopes often exhibit significant anisotropy, for example, when a beam is scattered into a grazing angle on surface interaction.
Proper wave tracing then requires support for full elliptical cone tracing, with accelerating data structure traversal and primitive (edges and triangles) intersection tests for arbitrary elliptical cones.
We provide implementation details and additional comparisons in our supplemental material.

\begin{figure*}[tb]
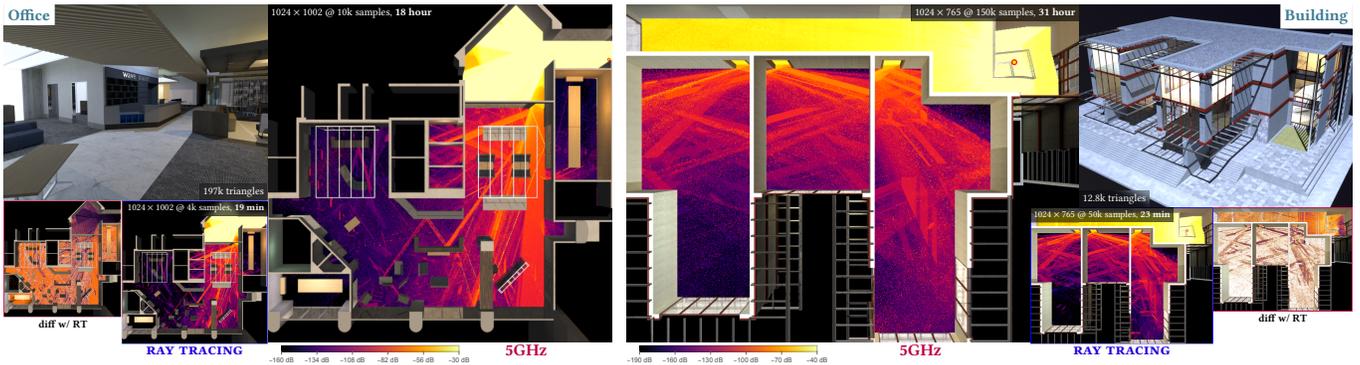

    \centering
    \begin{adjustbox}{width=1\linewidth}
        \inputtikz{results_indoor}
    \end{adjustbox}%
    \vspace*{-1.5mm}
    \caption{
        \textbf{Indoor 5GHz simulations.}
        We simulate signal coverage with a 5GHz wireless carrier in two indoor scenes with more challenging light transport.
        Visualized is the received signal strength (rss) in decibels that reaches the floor.
        No geometric simplifications are done; the geometry used for the wireless simulation is as visualized in the optical rendering insets.
        The surface materials are different from the optical materials: at 5GHz frequency, surfaces are effectively perfectly specular, and the material (refractive index) is selected from the appropriate construction material \cite{ITUR_P.2040_3}: wood, plywood, chipboard, plasterboard, concrete, or metal.
        (Office) This scene contains complex geometry and fine geometric details. The transmitter is placed at the top right corner; note the increased rss around the lobby on the left and the meeting room on the right.
        (Building) 
        rss is computed on the \ordinalnum{3} floor, while the transmitter is placed at \ordinalnum{2} floor level in the stairwell (top right).
        Due to the transmitter's placement, most emitted energy does not propagate to the \ordinalnum{3} floor, or is diffracted by the staircase or railing.
        Paths that diffract several times are needed to reach the far areas of the large room on the left.
        Both scenes contain many glass windows, through which significant energy radiates away.
        The elliptical conic envelopes' cross-section areas and primitive counts on interactions (means and standard deviations):
        (Office) \SI{1.1(1.364)}{\square\metre}, \num{34(470)};
        (Building) \SI{2(40)}{\square\metre}, \num{29(905)}.
        UTD-based free-space diffractions are used (see \cref{section_fsd}).
    }
    \label{fig:render_signal_coverage_indoor}
\end{figure*}%

\subsection{Results}   \label{sec:results}

We have validated our approach by comparing results to a BEM wave solver; see \cref{fig:wave_solver}.
We find that our renderer overall matches the reference well, yet computes solutions in 3D (rather than a 2D slice) while requiring a fraction in computation resources.
We have also evaluated our algorithms for optical rendering and modeling radio wave propagation.

In these results, we also report the means and standard deviations of the areas of the beams' cross-sectional areas and primitive count at interaction regions.
Note the large difference in areas between different wavelengths.
For convenience, these statistics are also summarised in \cref{table:performance} for all of our results.

\begin{table*}[!b]
\centering
\small
\begin{tblr}{
        colspec = { Q[2.2cm] r  Q[3.5cm,halign=r] Q[2.5cm,halign=r] Q[2.25cm,halign=r] }
    }
    \hline[2pt,blue1]
    \SetCell[c=2,r=2]{c} { \textsc{scene} }  &&
    \SetCell[c=2]{c} { \textsc{elliptical cone statistics} }  &&
    \SetCell[r=2]{c} { \textsc{throughput} \scriptsize (samples per millisecond) }
    \\
    \cline[dashed]{2-4}
    &&
    \SetCell{c} {\footnotesize cross-sectional area \\ \scriptsize (mean $\pm$ std. dev.)}  &
    \SetCell{c} {\footnotesize triangles per cone   \\ \scriptsize (mean $\pm$ std. dev.)}
    &
    \\
    \hline
    \SetCell[r=2]{l} \textbf{Etoile}   & \SI{30}{\giga\hertz}    &
    \SI{56(250)}{\square\metre} & \num{7(21)} &
    5.1k
    \\
     &  \SI{1}{\giga\hertz} &
    \SI{90(290)}{\square\metre} & \num{9(22)} &
    5.9k
    \\
    \SetCell[r=2]{l} \textbf{M\"unchen} &    \SI{30}{\giga\hertz}    &
    \SI{49(298)}{\square\metre} & \num{11(129)}  &
    900
    \\
     &  \SI{1}{\giga\hertz} &
    \SI{84(364)}{\square\metre} & \num{10(107)}  &
    1.1k
    \\
    \SetCell[r=1]{l} \textbf{Office}   & \SI{5}{\giga\hertz} &
    \SI{1.1(1.364)}{\square\metre} & \num{34(470)}  &
    150
    \\
    \SetCell[r=1]{l} \textbf{Building} & \SI{5}{\giga\hertz} &
    \SI{2(40)}{\square\metre} & \num{29(905)} &
    1.05k
    \\
    \SetCell[r=1]{l} \textbf{Room}     & optical &
    \SI{0.2(1.0)}{\square\milli\metre}  &   \num{14(261)}  &
    1.2k
    \\
    \hline[1pt,blue3]
\end{tblr}
\caption{
    \textbf{Performance statistics}.
    A summary of the average cross-sectional area of a traced elliptical cone on an interaction, as well as average count of triangles per interaction; as well as performance (samples per millisecond) for all of the scenes in our results.
}
\label{table:performance}
\end{table*}

\begin{figure}[t]
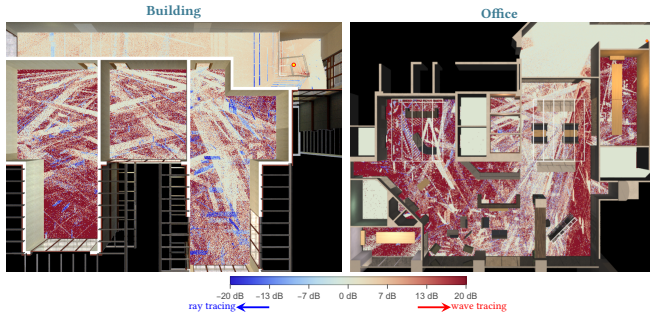

    \centering
    \begin{adjustbox}{width=1\linewidth}
        \inputtikz{rt_comparison}
    \end{adjustbox}%
    \vspace*{-1.5mm}
    \caption{
        \textbf{Received signal strength (rss) differences between ray tracing and wave tracing.}
        Blue indicates areas where a ray tracing simulation computes a higher rss compared with wave tracing, and red indicates regions where the situation is reversed.
    }
    \label{fig:rt_wt_diff}
\end{figure}%

\paragraph{Optical rendering}

Our optical results, \cref{fig:render_room}, demonstrate the ability of our implementation to simulate free-space diffractions in scenes with extensive geometric detail and complex light transport.
All renders are spectral and polarimetric.

These results were rendered with a bidirectional path tracing (BDPT) extension of our weakly-local path integral.
When forming connections between paths under BDPT, we wave trace each beam from the two path vertices involved in the connection strategy halfway towards each other.
If both beams are unoccluded, the beams meet and we integrate the beams over their respective space-frequency footprint \cite{Steinberg_rtplt}.
Compared with the classical approach, multiple importance sampling (MIS) for BDPT changes as some connection strategies may be occluded while others are not.
For our implementation, we use the classical strategy as an approximation, and leave correcting MIS for BDPT for future work.
See \cref{supp:subsection_bdpt} of our supplemental material for more information on our BDPT extension and explicit formulae.

\paragraph{Signal coverage}

We perform signal coverage simulations with long-wavelength radiation in two large-scale city scenes (\cref{fig:render_signal_coverage}) and two more complex indoor scenes (\cref{fig:render_signal_coverage_indoor}).
At these wavelengths, all the surfaces become perfectly specular, and light transport is simulated unidirectionally from the emitter only.
The wavelength-dependent complex refractive indices for all materials are given by \citet{ITUR_P.2040_3} recommendations for construction materials, and include concrete, glass, brick, wood, and metal surfaces.
For comparison, same scenes are rendered with pure ray tracing; note the significantly higher received signal strength that reaches the shadowed regions with wave tracing.
See our supplemental for additional analysis.


\subsubsection{Additional Comparison with ray tracing}

We compare the received signal strength (rss) that reaches the floor in the Building and Office scenes between ray tracing and wave tracing, see \cref{fig:rt_wt_diff}.
The color-coded blue and red areas indicate areas where ray tracing or wave tracing, respectively, produce a greater rss.
As expected, over most areas wave tracing gives rise to a higher rss, as light is able to diffract and penetrate into harder-to-reach regions.
Nevertheless, in some areas ray tracing dominates, at times substantially so.
These blue areas are \emph{errors induced by the ray tracing simulation}:
light that should have diffracted continues to propagate unobstructed.
For example, light that passes through the railings around the stairwell in the Building scene (top right), or light that is reflected by ceiling fixtures towards the lobby in the Office scene (left).


\subsubsection{Comparison with Sionna}

\emph{Sionna} \cite{sionna} (version 0.19) is a communications systems simulation framework that employs UTD to simulate diffractions.
An initial ray tracing pass is used to find triangles and their edges that might participate in diffraction.
All these edges, across all samples, are stored in memory, and a subsequent pass attempts to connect emitter--edge--receiver diffraction paths via every found edge and to every receiver (or pixel in a coverage map).
No diffractions are mixed with reflections or scattering events, likewise no multiple diffractions, or multi-edge diffractions are simulated.

\begin{figure}[t]
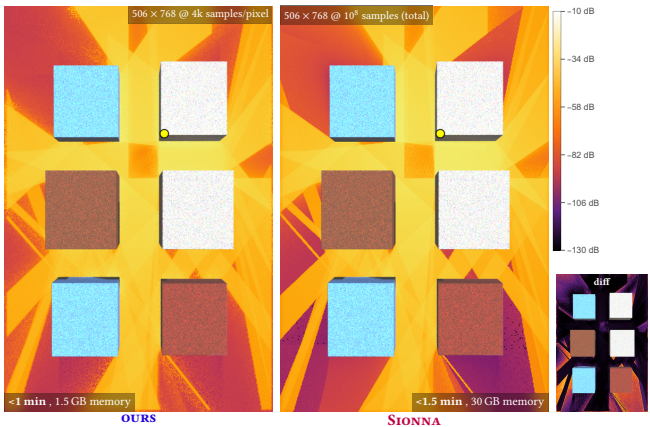

    \centering
    \begin{adjustbox}{width=1\linewidth}
        \inputtikz{sionna}
    \end{adjustbox}%
    \vspace*{-1.5mm}
    \caption{
        \textbf{Comparison with Sionna.} 
        The scene is ``simple street canyon'' from \citet{sionna}.
        The position of emitter illustrated by the yellow circle.
        The received signal strength (rss) in decibels that reaches the street level is visualized, with a rendering of the occluding and diffracting building geometry overlaid (buildings' colors are for visualization only).
        Not only can our approach account for path that involve both diffractions and reflections, but ours uses significantly less memory as well.
    }
    \label{fig:sionna}
\end{figure}

We do a comparison with Sionna in a very simple scene, where single-edge, direct (emitter--edge--receiver) diffractions dominate, see \cref{fig:sionna}.
There are small differences between the material and emitter models, nevertheless most differences are minor.
Most differences arise in the bottom parts of the image, where the majority of the energy arrives by reflecting off the buildings first and then diffracting, which cannot be simulated with Sionna.

Note the memory requirements:
Sionna's memory requirements become prohibitive with more complex scenes that require higher sample counts.

\section{Conclusion}  \label{section_conclusion}

Our discussion began with analyzing the power---in terms of reproducing wave-optical phenomena---of the classical path integral formulation of light transport; then, we generalized it to two formulations that are used in wave simulations:
(i) ray-based bilinear transport; and (ii) weakly-local (region-to-region) transport.
Region-to-region path tracing, as formalized by our weakly-local path integral formulation, enables formulating---and in turn, importance sampling---scattering functions that account for the entire interfering interaction within a region.
Contrast this to typical shooting-bouncing rays or UTD-based approaches, where interference is resolved later at the sensor (as formalized by our bilinear path integral), and importance sampling turns into a global problem.

We discussed how to design a system for wave simulations, based on our weakly-local path integral formulation, and we release a complete rendering system that implements the ideas in this paper.
Our renderer is a fully polarimetric spectral renderer, designed from scratch to target elliptical cone tracing and wave simulations across the EM spectrum.
Our system supports multi-material and multi-edge interactions, as well as multiple such interactions over a path.

Both our path integral generalizations serve to relax some of the limitations of the classical path integral, discussed in \cref{sec:classical_path_integral}.
The far-field and high-frequency assumptions are often relaxed (the degree of which depends on the optical formalism), and the restriction to statistically stationary materials is eliminated entirely.
Future work may focus on eliminating the time-averaging assumption, enabling interference across spectral samples for acoustics simulations, for example.
Future work may also target the differentiability of our rendering system for wave-optics modeling; differentiable rendering is likely to be more efficient in a non-bilinear framework like ours.


\bibliographystyle{ACM-Reference-Format}
\bibliography{paper}

\end{document}